\documentclass[letterpaper, conference, 10 pt]{ieeeconf}  
\usepackage[left=50pt,right=50pt,top=60pt,bottom=43pt]{geometry}
\IEEEoverridecommandlockouts                              
\usepackage{graphics} 
\usepackage{epsfig} 
\usepackage{mathptmx} 
\usepackage{times} 
\usepackage{amsmath} 
\usepackage{amssymb}  
\usepackage{booktabs} 
\usepackage{cite}
\usepackage{amsfonts}
\usepackage{amsbsy}
\usepackage{multirow}
\usepackage{makecell}

\title{\LARGE \bf
Investigation on a Novel Length-Based Local \\ Linear Subdivision Strategy for Triangular Meshes
}

\author{Junyi Shen$^{1}$
\thanks{* This work is not supported by any organization.}
\thanks{$^{1}$Department of Information Physics and Computing, Graduate School of Information Science and Technology, The University of Tokyo, 7-3-1 Hongo, Bunkyo-Ku, Tokyo, Japan
        {\tt\small junyi-shen@g.ecc.u-tokyo.ac.jp}}
}

\begin{document}

\maketitle
\thispagestyle{empty}
\pagestyle{empty}

\begin{abstract}

Triangular meshes are a widely used representation in the field of 3D modeling. In this paper, we present a novel approach for edge length-based linear subdivision on triangular meshes, along with two auxiliary techniques. We conduct a comprehensive comparison of different subdivision methods in terms of computational capabilities and mesh-enhancing abilities. Our proposed approach demonstrates improved computational efficiency and generates fewer elements with higher quality compared to existing methods. The improvement in computational efficiency and mesh augmentation capability of our method is further enhanced when working with the two auxiliary techniques presented in this paper. Our novel strategy represents a significant contribution to the field and has important implications for local mesh refinement, computer-aided design, and isotropic remeshing.

\end{abstract}

\section{INTRODUCTION}

Mesh generation has gained considerable attention across various disciplines, including computer graphics \cite{zhang2010spectral, sorkine2006differential}, numerical analysis \cite{he2014review, sorkine2006differential}, automatic path planning \cite{petres2007path, bircher2018receding}, and intelligent manufacturing \cite{raj2000modeling, zhao2020mixed}. It has proven to be a valuable modeling tool and is now included in almost all standard modeling packages, such as 3ds Max, Solidworks, and Maya \cite{bolz2002rapid}. In recent decades, numerous tessellation methods for meshing analytical 3D surfaces using geometric primitives such as triangles, quadrilaterals, and general n-sided polygons have become available \cite{guo2019automatic}. Among all types of tessellations, the triangular surface mesh is the most popular for manufacturing and visualization because of its exceptional ability to represent complex free-form surfaces \cite{alliez2003isotropic}, ensure numerical accuracy, and offer reshaping flexibility \cite{wu1997advantages, lindquist2005comparison}. 

Mesh size and isotropy play a critical role in numerical simulation and analysis by determining computational accuracy and the amount of computation required  \cite{he2014review, jain2017numerical}. In numerical simulations and analysis, smaller element sizes lead to greater accuracy but require longer computation time. Coarse meshes can be refined or subdivided to reduce the error introduced by modeling a free-form surface into discrete facets \cite{sosnowski2018polyhedral, chen2002automated}. The geometric property like isotropy and distortion of the mesh significantly impacts the performance of finite element analysis computations since it affects both the approximation error of the finite element solution and the spectrum of the associated stiffness matrix \cite{du2009mesh}. With the growing integration of numerical simulation and analysis into the manufacturing process, the significance of mesh size and distortion is increasingly apparent, as companies rely on performance evaluations based on simulation results earlier in the design stage \cite{beall2003accessing}. 

A high-quality mesh strikes a balance between simulation accuracy and computational cost. In contrast, low-quality meshes characterized by extreme thinness or distortion impede solution convergence, increase analysis errors, and the computational burden. Geometric characteristics, including shape, angle, and complexity \cite{bern1995mesh}, directly impacting the isotropy and distortion of meshes, which further affects the quality of the mesh group. Finding the optimal balance between simulation precision and computational cost requires careful consideration of multiple factors related to mesh size, distortion, and distribution. Although most commercial mesh generators support mesh size control, they typically require manual intervention from engineers to modify meshes to achieve the desired isotropy and uniform distribution \cite{shimada2011current}. 

\begin{figure}[tb]
    \centering
\includegraphics[width=\linewidth]{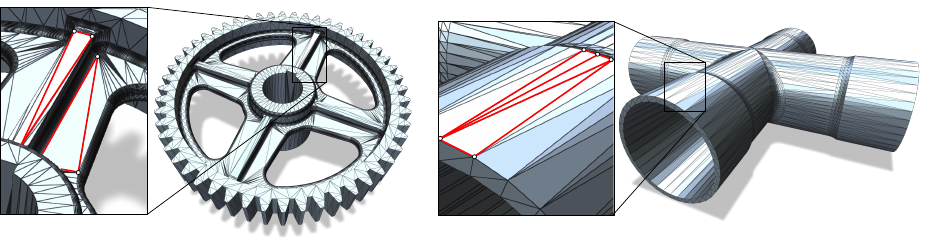}
    \caption{Examples of the non-uniformly distributed and highly stretched meshes of models generated by commercial CAD software.}
    \label{fig:1.1}
\end{figure}

The use of computer aided design (CAD) software for generating meshes has become prevalent across different fields, including finite element analysis (FEA) \cite{foucault2008adaptation}, rapid prototyping \cite{jamieson1995direct, ledalla2018performance}, computational fluid dynamics (CFD) \cite{king2006cad}, and visualization \cite{flynn1989cad}. There are packages such as NetGen \cite{schoberl1997netgen} and Gmsh \cite{geuzaine2009gmsh} providing the capability to represent 3D models in a discrete manner. However, meshes generated by commercial CAD software are often coarse and non-uniform, with flat or nearly flat sections coarsely represented by larger triangles due to the lesser complexity and continuity of free-form surfaces in industrial workpiece models. Highly curved sections, on the other hand, are finely represented by a great number of small triangles resulting from the preservation of local minute details \cite{guo2019automatic, sunil2008automatic}. Fig. \ref{fig:1.1} depicts two examples of non-uniformly distributed and highly stretch coarse meshes generated by commercial CAD softwares. 

Apart from CAD, the refinement of surfaces is also a crucial aspect in the application of computing operations between surfaces and geometric design \cite{severn2006fast}. The efficient intersection of facets makes it possible to apply Boolean operations on meshes, creating new objects from existing ones. However, to produce high-quality results, intersections need to be performed on meshes with high resolution. The computation of intersections on the hierarchy of meshes resulting from the subdivision is critical and cannot be efficiently solved for meshes with low resolution using existing mesh refinement strategies.

The issue of low mesh quality in numerical simulation and investigation has been the subject of ongoing research. Various enhancement techniques, such as mesh optimization, remeshing, and mesh refinement, have been developed to address this problem \cite{hoppe1993mesh, cignoni1998comparison, alliez2008recent}. To meet the specific requirements of different numerical simulation and analysis, adaptive processing is essential \cite{beall2003accessing}. Mesh refinement strategies based on specific standard, such as edge length limit, have been proposed to achieve this goal. For example, Noii et al.\cite{noii2020adaptive} demonstrated the effectiveness of refining meshes using edge length as a criterion for phase-field fracture problems. Meanwhile, Guo et al.\cite{guo2019automatic} described a mesh optimization strategy that utilizes user-specified edge length.

This study specifically investigates the mesh subdivision process, a mesh refinement technique used to address the challenges associated with low-resolution meshes and defective geometric properties. We determine the need for mesh subdivision based on the edge length.

\subsection{Related Works}

Over the past few decades, several subdivision methods have been proposed, including the Catmull-Clark method \cite{catmull1998recursively}, modified butterfly scheme \cite{rypl2006generation}, and Loop technique \cite{loop1987smooth}, have appeared and been developed over the years. However, these methods have been deemed unsuitable for manufacturing purposes as they can distort the original shape and produce non-manifold faces \cite{ledalla2018performance}. 

Linear subdivision, which involves connecting the midpoints of three edges in a triangular mesh and generating four new triangles, has been demonstrated to refine discretely represented models while perfectly preserving their geometric structure. This technique has been used in generating point cloud models \cite{jacquemin2023smart} and as a pre-processing step in other isotropic remeshing operations \cite{surazhsky2003isotropic}. Linear subdivision can be performed globally on the entire mesh group or locally on specific meshes based on user-defined criteria such as triangle shape, edge length, and location. Packages like PyVista \cite{sullivan2019pyvista} can efficiently perform global linear subdivisions by processing with arrays and matrices. However, the global subdivision requires the input surface to be manifold and generates numerous redundant triangles by indiscriminately dividing every mesh. Additionally, global processing can only operate based on the specified number of subdivide operations, to meet the user-specific requirement an additional checking is necessary before or after the subdivision operation, which complicates the process. In contrast, local subdivision has higher degrees of freedom. It can be performed automatically on any discrete mesh without restrictions while reducing the number of new meshes generated and, subsequently, alleviating the computational burden in the subsequent computing process. However, local usually subdivision incurs higher computational expense than global subdivision.

In the field of computer-aided design (CAD) modeling, the current linear local subdivision method that connects all the midpoints of three edges has limitations that need to be addressed. Precisely,  when highly stretched triangles are present in the input model, some edges within the user-specified limit may still be cut in half, resulting in the generation of redundant vertices and meshes. Moreover, the produced segmentations inherit parent poor geometric properties, such as excessive distortion and high anisotropy, from the unprocessed parent mesh, leading to an increase in the proportion of meshes with suboptimal qualities for further processing. Although previous research has explored strategies to improve the performance of global or local linear subdivisions, such as the Green rule \cite{bank1983some} and bisection \cite{bern2000mesh}, there is still a vacuum in the corresponding relationship between the specified shape of triangles and the subdividing strategy.

This study proposes a novel method to address the issues associated with using the classic local subdivision in coarse mesh models generated from CAD packages. The new approach differentiates the division scheme for various triangle types based on the number of unqualified long edges, offering several advantages such as efficient performance, automatic adaptive processing, and improved mesh geometry suitable for manufacturing purposes. In addition, we introduce two auxiliary techniques that can be integrated with our proposed method to further enhance its performance. A comprehensive comparison is conducted to demonstrate the effectiveness of our approaches. Finally, we provide conclusions.

\section{Proposed Subdivision Strategy} 

\begin{figure}[tb]
    \centering
\includegraphics[width=\linewidth]{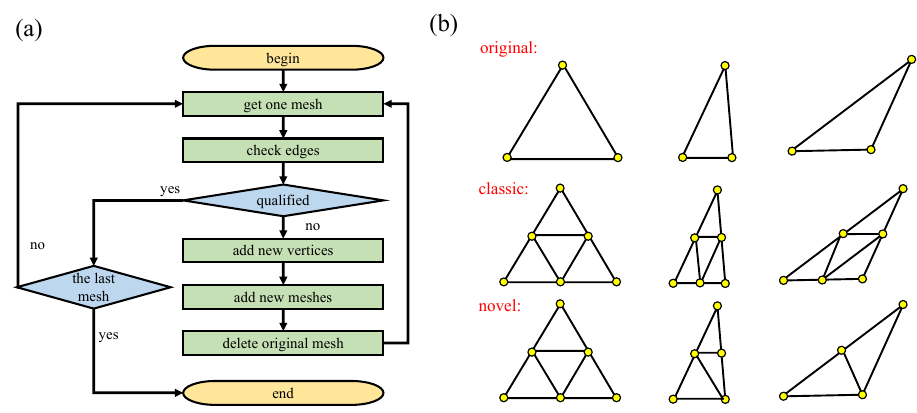}
    \caption{(a) Flowchart of the recursive operation employed in local subdivisions, (b) Different refinement schemes corresponding to triangles with different numbers of qualified edges.}
    \label{fig:2.1}
\end{figure}

The most straightforward approach for local subdivision is to use a recursive algorithm that involves repetitively applying subdivision stencils to achieve the desired mesh lengths. This approach provides adaptive rendering and multiresolution surfaces with excellent performance, although optimal results require significant effort \cite{muller2000subdivision}. In our study, we also implement local subdivisions using a recursive algorithm, as shown in Fig. \ref{fig:2.1}. By recursively removing and adding meshes onto a stack list, our method ensures a thorough search and operation on the input mesh group. Alternatively, local subdivisions can be implemented in a depth-first manner, which requires a smaller memory footprint and has the potential for hardware implementation \cite{pulli1996fast}. 

The classic technique for local subdivision mandates that any mesh with edges exceeding the user-specified threshold should be quartered into four smaller ones with the same shape in a single operation. This gives rise to a relationship between the number of original meshes, the increased meshes, the subdivision operations, and the user’s edge length demand, which can be expressed as 
\begin{equation}
N = \sum^n_{i=0}\sum^{k_i}_{u=0} 4^u,
\label{N}
\end{equation}
\begin{equation}
k_i = \lceil \log_2 (\max(L_i)/L_{\text{threshold}}) \rceil.
\label{k_i}
\end{equation}
Here, $N$ represents the total number of meshes resulting from subdividing an original mesh group containing $n$ meshes, with the largest edge length being $ \max(L_i) $, and individual mesh must be subdivided for $k_i$ times to meet the user's desired edge length threshold $L_{\text{threshold}}$ .

In contrast to the classic method, our proposed strategy differentiates among division schemes for triangular meshes based on the number of edges falling within the user-specified limit. Our approach ensures that no edges below the limit being cut, identifying three distinct subdivision types for triangular meshes with zero, one, and two qualified edges (whose length dose not exceed the user-specified limit), respectively. Fig. \ref{fig:2.1}(b) illustrates the refinement of triangular meshes with varying numbers of qualified edges. For meshes with three long unqualified edges, our proposed strategy aligns with the classic approach by connecting the midpoints of three edges. For meshes with two unqualified edges, as shown in the middle part of the figure, our method connects the midpoints of the two relatively longer sides and joins the center of the longest edge to its opposite point. For meshes with only one unqualified edge, as shown at the bottom of the figure, we apply a bisection by connecting the midpoint of the unqualified edge to its opposite point. As described above, our approach divides each unqualified coarse mesh into two, three, or four segments instead of all four smaller ones, thus preventing the generation of redundant elements and the inheritance of the original structure by the generated meshes.

\begin{figure}[tb]
    \centering
\includegraphics[width=\linewidth]{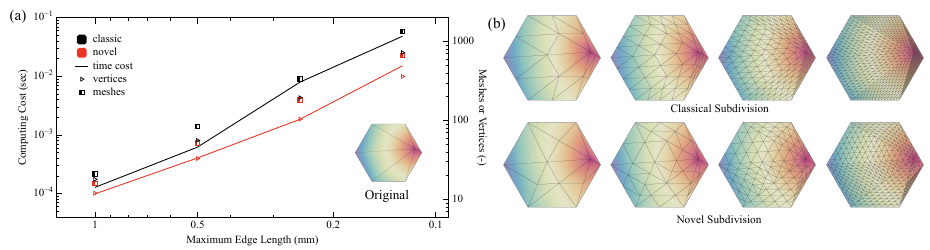}
    \caption{(a) Computational time cost and the quantities of generated vertices or meshes as versus the edge length limit, (b) Subdivision process of triangular meshes using both methods.}
    \label{fig:2.2}
\end{figure}

Fig. \ref{fig:2.2} provides a visual comparison between the classic and proposed methods for subdividing a hexagon consisting of six triangular meshes. Panel (a) exhibits the exponential growth rate in computational cost and elements generation as the limit for edge length of triangle meshes shrinks, while (b) highlights the geometric properties of meshes produced by different approaches. The results indicate that our proposed method is more efficient and effective than the classic approach, requiring less time and generating fewer vertices and meshes to achieve the user-specified length requirement. This reduction in the number of vertices and faces leads to a reduction in the computational cost of subsequent geometry processing. All the programs in this work were run on an Apple Silicon M1 with 8 cores and 8GB RAM. 

\section{Auxiliary Techniques}

In some cases, large triangular meshes in 3D models with three long edges may have defective geometric properties, such as extreme thinness or excessive stretching (as depicted in Fig. \ref{fig:1.1}). Subdividing such meshes using the classic technique results in the generated triangles to be easily unqualified for more stringent user demands \cite{sunil2008automatic}. Since our primary strategy for meshes with three long edges aligns with the classic method, it needs to address the issue of preserving the original degenerated geometric structure, or may cause generated meshes with poor quality. To tackle this issue, we propose two auxiliary techniques that aim to break the original poor geometric structure and thereby enhance the quality of the generated meshes, further improving the mesh-enhancing capacity of our new approach.

\subsection{Multistage Novel Subdivision}

\begin{figure}[tb]
    \centering
\includegraphics[width=\linewidth]{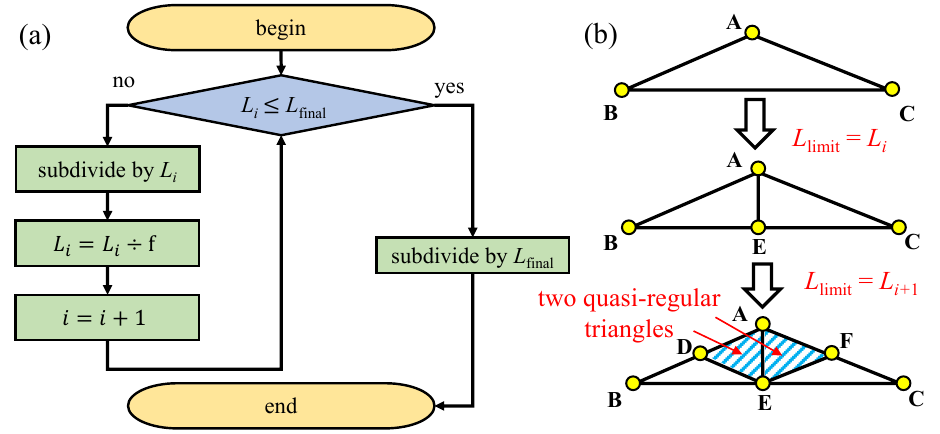}
    \caption{Diagrammatic of the multistage subdivision.}
    \label{fig:3.1}
\end{figure}

The concept of the multistage novel subdivision is illustrated in Fig. \ref{fig:3.1}, where (a) illustrates the decreasing of edge length limit, and (b) provides an example of how the multistage subdivision breaks the original mesh structure and generates meshes with high qualities. The multistage subdivision takes an initial length limit $L_0$, which is typically high enough to ensure adequate subdivide stages, a fold factor $f$, and a final edge length limit $L_\text{Final}$ as input. The process is carried out continuously on the meshes generated in the previous stage, with each successive operation reducing the upper limit on edge lengths by a divisor (fold factor) $f$. The edge length limit after $k$ stages is given by $L_k=L_0/f^k ,(k=0,1,2,…)$.

The multistage subdivision is initiated with an edge length limit exceeding the desired value. This is designed to avoid the regular subdivision in the early stages, resulting in two or three, more structurally optimal meshes with different shapes instead of four identical ones (as shown in Fig. \ref{fig:2.1}). The process continues with subsequent subdivisions using more stringent length limits until the user's requirements in edge length are met.

Incorporating our novel subdivision method into the multistage subdivision process offers significant advantages for triangulated 3D models. By repeatedly applying the novel subdivision, we address initial poor geometric properties and generate discrete representations with fewer elements produced. Furthermore, the multistage subdivision is a simple yet effective approach that reduces computational requirements and processing needs. It is worth noting that the improved performance achieved by integrating the multistage subdivision strategy is a direct result of the unique ability of our proposed method to modify the geometric properties of triangular meshes. Applying this multistage concept to traditional subdivision techniques would not yield similar results.

\subsection{Angle-Restricted Novel Subdivision}

\begin{figure}[tb]
    \centering
\includegraphics[width=\linewidth]{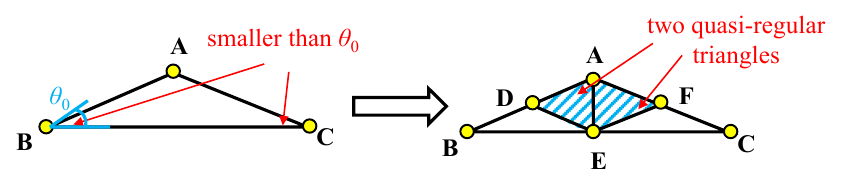}
    \caption{The angle-restricted subdivision and the elementary subdivision.}
    \label{fig:3.2}
\end{figure}

The angle-restricted technique, presented in Fig. \ref{fig:3.2}, is designed explicitly for meshes containing edges that are all longer than the length limit and two acute angles smaller than a predefined angle threshold $\theta_0$. 

Unlike the primary novel method, which connects the midpoints of the two sides that enclose the largest angle, the angle-restricted subdivision connects those midpoints to the midpoint of the longest edge. By utilizing this approach, the original mesh's poor geometric properties are inherited by only two of the four generated triangles, thereby disrupting the geometric structure of poorly shaped meshes with three long edges and increasing the proportion of well-shaped meshes.

\section{Results and Discussion}

In this section, we present a comprehensive comparison between our proposed strategies and the classic method. The evaluation is based on various metrics, including the anisotropy and density distribution of the generated meshes, element generations, computational expenses, and mesh-enhancing capacities. To compare the results of the multistage subdivision with the elementary novel subdivision, we select fold factor $f$ values of 1.5, 1.8, and 2.0. Moreover, we compare the performance of the angle-restricted subdivision with different threshold angle $\theta_0$ values of 15°, 30°, and 45° to that of the primary novel approach.

\subsection{Distribution and Anisotropy}

\begin{figure}[tb]
    \centering
\includegraphics[width=\linewidth]{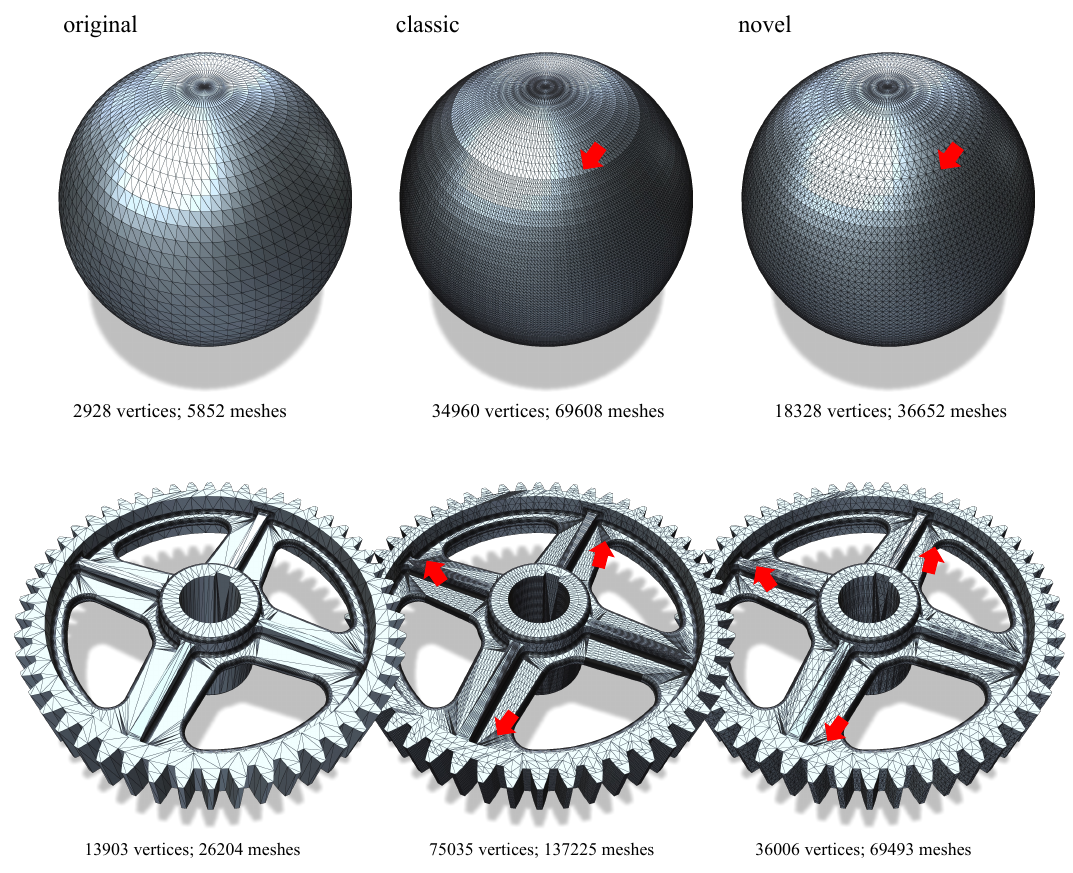}
    \caption{Subdivision results using different methods.}
    \label{fig:4.1}
\end{figure}

Fig. \ref{fig:4.1} presents the results of subdividing models with simple (sphere) and complex (gear) geometric structures. Our proposed approach generates a more uniform mesh distribution with fewer redundant elements than the classic method (as indicated by red arrows). The new strategy achieves a better distribution when performed on simple geometric structures with a limited variation of input mesh shapes. However, our method can still effectively address the defective features when subdivided by the classic approach, even for complex original geometric structures.

To better observe and evaluate processed meshes' anisotropy and density, we introduce an assessment factor defined by 
\begin{equation}
b_i=  \min(L_i)/L_\text{limit}, (i=1,2,\dots,M).
\label{b_i}
\end{equation}
Here $b_i$ represents the ratio of the shortest length of any edge containing the vertex $b_i$, $L_\text{limit}$ is the current edge length limit, and $M$ is the total number of vertices.

The value of $b$ ranges from 0 to 1 for each vertex. A mesh group with more vertices having the $b$ value closer to 1 indicates better anisotropy and more uniform density, corresponding to fewer redundant meshes. Conversely, a mesh group with more vertices having the $b$ value closer to 0 indicates poorer anisotropy and non-uniform density, with numerous redundant elements produced.

\begin{figure}[tb]
    \centering
\includegraphics[width=\linewidth]{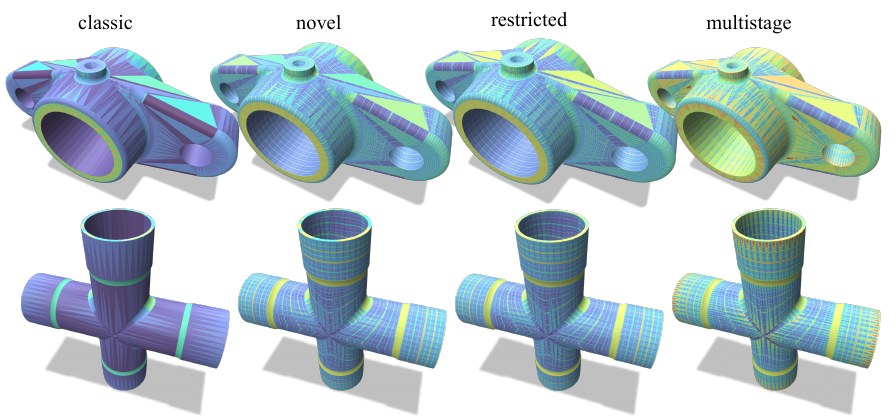}
    \caption{Distribution of $b$ values among meshes produced by four approaches. The color scale indicates that the darker the blue, the closer the value $b$ is to 0, and the darker the red, the closer the value $b$ is to 1.}
    \label{fig:4.2}
\end{figure}

The distributions of $b$ for two models with simple (cylinder) and complex (bearing) geometric structures, subdivided by four different schemes, are presented in Fig. \ref{fig:4.2}. The results from the classic approach show clear boundaries and patches among meshes due to the variability in the geometric properties of the unprocessed input meshes. In contrast, although our novel method still displays some borders between mesh clusters, it disrupts those patches with good quality meshes and exhibits generally better isotropy and more uniform density. The angle-restricted method shows differences that are not significantly observable compared with the elementary novel method. However, its ability to handle poorly shaped triangles is still illustrated in some areas. The multistage subdivision performs best in this comparison, effectively addressing input meshes' initially flawed geometric characteristics compared with the other three approaches. 

\subsection{Computational Expense and Elements Generation}

\begin{figure}[tb]
    \centering
\includegraphics[width=\linewidth]{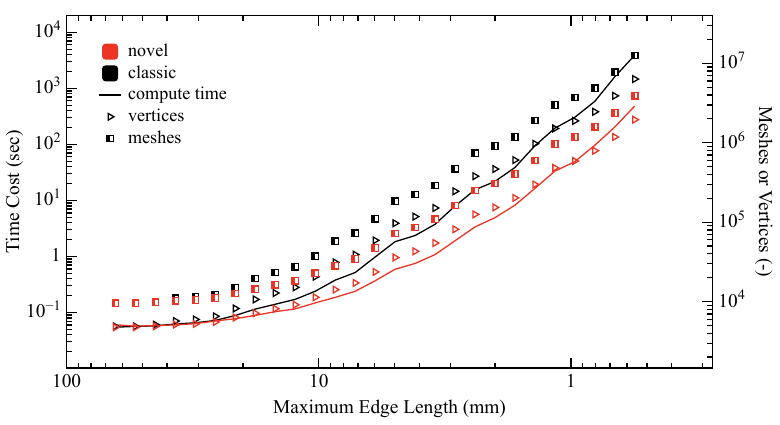}
    \caption{Computing time and the elements generated versus the maximum edge length.}
    \label{fig:4.3}
\end{figure}

The result illustrated in Fig. \ref{fig:4.3} highlights the increase in computing costs and the generation of elements as the limit of edge length decreases. Most unprocessed coarse meshes contain three qualified edges, with only a limited number requiring subdivision, making it difficult to differentiate between the two methods when the length limit is relatively high. With the maximum edge length decreasing, both the classic and novel methods experience an exponential increase in computing time and element generation. However, our novel method tends to increase by a slower rate. It is expected that as the threshold edge length continues to decrease, the resource-saving benefits of the novel strategy will become increasingly evident and valuable. 

The proposed method's efficient subdivision capability arises from its ability to identify qualified edges in the triangles that require processing (as illustrated in Fig. \ref{fig:1.1}). This leads to a reduction in the number of vertices and faces and a decrease in computing time and memory usage not only during the subdivision process but also during any subsequent geometric processing. In contrast, the classic method lacks this capability and generates more vertices and faces, resulting in higher computational expenses.

\begin{figure}[tb]
    \centering
\includegraphics[width=\linewidth]{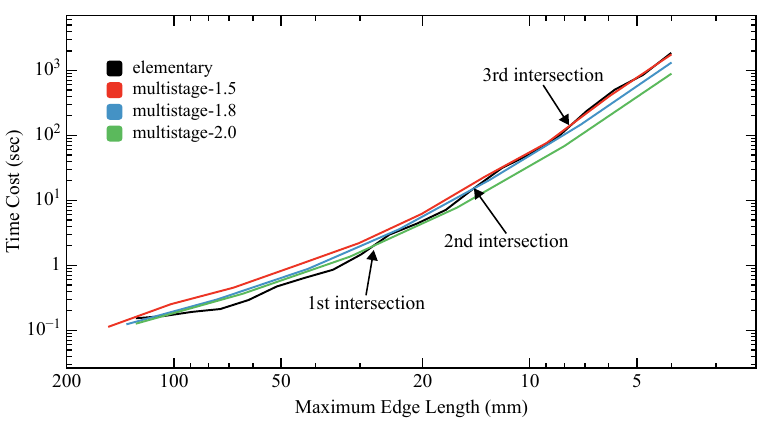}
    \caption{Comparison in computing time of different subdivision strategies, the multistage subdivision with the fold factor $f$ being 2.0 shows the least time-consuming in the initial stages and the fastest convergence (1st intersection) to the elementary novel method, followed by the value of $f$ being 1.8 and 1.5.}
    \label{fig:4.4}
\end{figure}

Fig. \ref{fig:4.4} presents a comparison of the time cost between the multistage and elementary novel subdivisions. The results indicate that when the edge length limit is relatively high, the multistage subdivision consumes more computational time than the elementary method. This is attributed to the cumulative effect of multiple operations performed in each stage. However, the computational time of the multistage strategy shows a slower growth rate compared to the elementary one. Eventually, when the time cost incurred by the initial stages becomes smaller than the time saved due to the slower growth rate, the advantage of the multistage subdivision in terms of computational time becomes evident. Additionally, in the scope of this work, as the fold factor $f$ increases, the computational time incurred in the initial stages decreases, and the multistage subdivision has a faster convergence toward the elementary novel method. This is because the larger the $f$, the faster the length limit decreases, and the fewer substages are applied.

\begin{figure}[tb]
    \centering
\includegraphics[width=\linewidth]{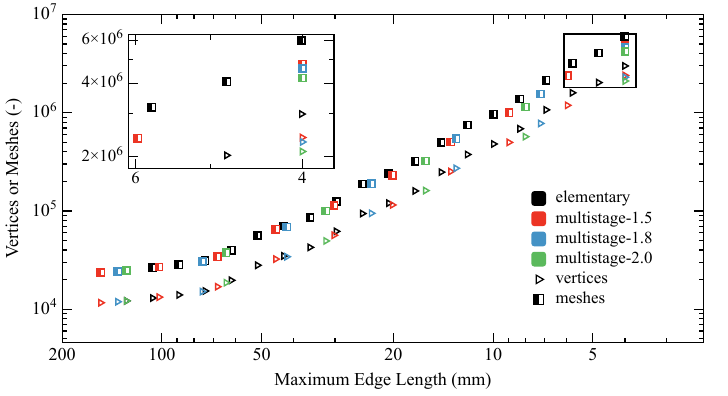}
    \caption{Comparison in the number of generated elements using multistage subdivisions with different parameters and that using the elementary novel subdivision.}
    \label{fig:4.5}
\end{figure}

The results in Fig. \ref{fig:4.5} demonstrate that the growth rate of vertices and meshes generated by multistage subdivisions is slower than that of the elementary novel subdivision as the maximum edge length decreases. It is worth noting that among the multistage methods, the larger the fold factor $f$, the fewer the number of vertices and meshes generated during the subdivision process.

\begin{figure}[tb]
    \centering
\includegraphics[width=\linewidth]{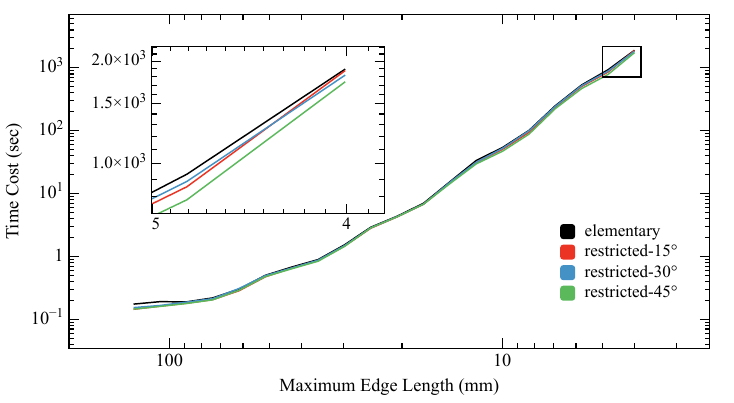}
    \caption{Comparison in the computing time using angle-restricted subdivisions with different parameters and that using the elementary novel subdivision.}
    \label{fig:4.6}
\end{figure}

\begin{figure}[tb]
    \centering
\includegraphics[width=\linewidth]{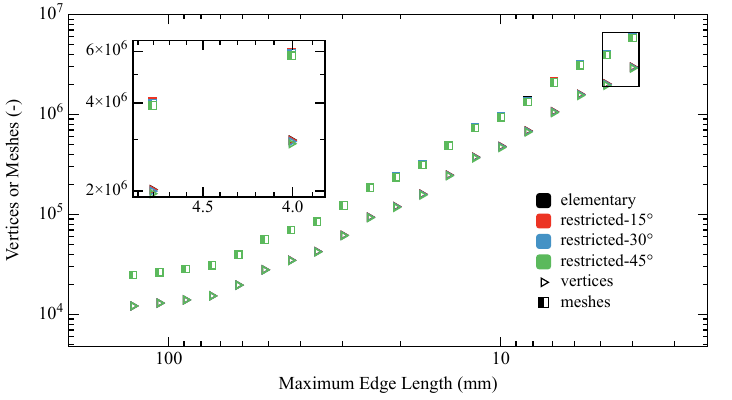}
    \caption{Comparison in the number of generated elements using angle-restricted subdivisions with different parameters and that using the elementary novel subdivision.}
    \label{fig:4.7}
\end{figure}

The results of the angle-restricted subdivision are presented in Fig. \ref{fig:4.6} and Fig. \ref{fig:4.7}, which show the analysis of the computational time and element generation, respectively, compared to the elementary subdivision. The results indicate that although the advantage brought by the angle-restricted method is not immediately observable, as the edge length limit decreases, the reduction in computing time and element generation becomes increasingly noticeable. Table \ref{tab} provides a more direct comparison of the results from different methods. The data show that the larger the threshold angle $\theta_0$, the more time is saved, and the fewer elements are generated during the subdivision process.

\begin{table*}[h]
\centering
\caption{Subdivision result of different methods}
\label{tab}
\begin{tabular}{|c|c|c|c|c|c|c|c|c|c|c|c|c|}
\hline
\multirow{2}{*}{\textbf{approach}} & 
\multirow{2}{*}{\begin{tabular}[b]{@{}c@{}}\textbf{vertices} \\ \(\boldsymbol{(10^3)}\)\end{tabular}} & 
\multirow{2}{*}{\begin{tabular}[b]{@{}c@{}}\textbf{meshes} \\ \(\boldsymbol{(10^3)}\)\end{tabular}} & 
\multicolumn{4}{c|}{\textbf{angle}} & \multicolumn{3}{c|}{\textbf{quality (q)}} & \multirow{2}{*}{\textbf{time (sec)}} \\
\cline{4-10}
& & & \textbf{\textless 15°} & \textbf{\textless 30°} & \textbf{\textgreater 90°} & \textbf{\textgreater 120°} & \textbf{\textless 0.3} & \textbf{\textless 0.5} & \textbf{\textgreater 0.8} & \\
\hline
classic & 5541 & 11026 & 12.4\% & 27.0\% & 22.0\% & 2.3\% & 26.6\% & 46.1\% & 10.8\% & 3900 \\
\hline
elementary novel & 2991 & 5985 & 4.7\% & 17.6\% & 13.4\% & 1.8\% & 8.6\% & 22.4\% & 37.2\% & 1896 \\
\hline
restricted (15°) & 2982 & 5968 & 4.5\% & 17.4\% & 13.3\% & 1.7\% & 8.2\% & 22.1\% & 37.3\% & 1877 \\
\hline
restricted (30°) & 2952 & 5908 & 4.5\% & 16.9\% & 13.1\% & 1.5\% & 8.0\% & 21.5\% & 38.3\% & 1820 \\
\hline
restricted (45°) & 2903 & 5810 & 4.5\% & 16.8\% & 12.5\% & 1.4\% & 7.9\% & 21.0\% & 39.6\% & 1740 \\
\hline
multistage (1.5) & 2396 & 4796 & 1.2\% & 5.1\% & 9.8\% & \textbf{0.1\%} & \textbf{2.8\%} & \textbf{7.7\%} & 60.7\% & 1767 \\
\hline
multistage (1.8) & 2298 & 4599 & \textbf{1.1\%} & \textbf{4.4\%} & \textbf{7.8\%} & 0.2\% & 2.9\% & 7.9\% & \textbf{65.6\%} & 1324 \\
\hline
multistage (2.0) & \textbf{2101} & \textbf{4205} & 2.2\% & 9.2\% & 9.0\% & 1.9\% & 5.9\% & 13.3\% & 64.7\% & \textbf{892} \\
\hline
\end{tabular}
\end{table*}

\subsection{Mesh Qualities and Angles}

\begin{figure}[tb]
    \centering
\includegraphics[width=\linewidth]{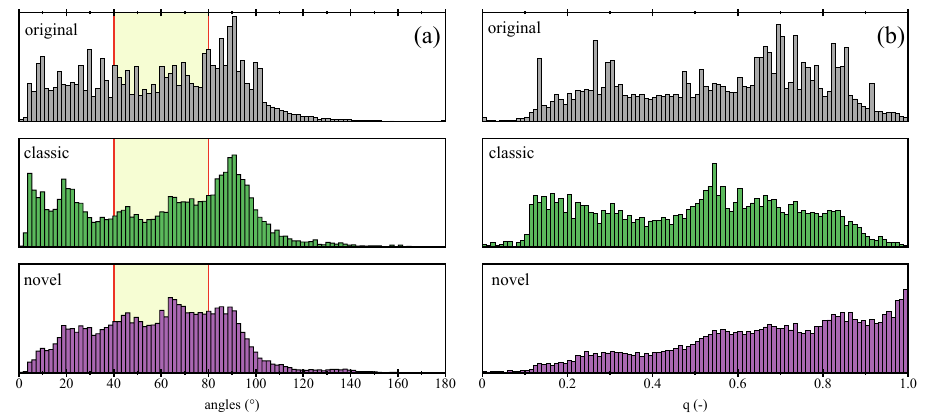}
    \caption{the comparison in angle distributions, where the highlighted part is the ideal angle range (40° to 80°), and (b): the comparison in mesh quality distributions.}
    \label{fig:4.8}
\end{figure}

Fig. \ref{fig:4.8} provides an illustration of the disparities in the angle and quality distributions, respectively. As depicted, the classic subdivision results in meshes that inherit the suboptimal geometric properties of unprocessed meshes and amplifies the proportion of poor geometric properties. In contrast, our novel method produces a distinct angle distribution resulting from the elimination of the initial poor geometric structure in highly anisotropic meshes. The novel method raises the proportion of angles within the ideal range of 40° to 80°, while decreasing the proportion of extraordinarily acute or quasi-right angles, thereby improving the quality of the original mesh groups \cite{guo2019automatic}.

\begin{figure}[tb]
    \centering
\includegraphics[width=\linewidth]{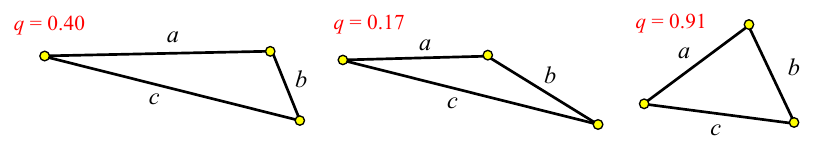}
    \caption{Examples of the $q$ value in some typical triangle types.}
    \label{fig:4.9}
\end{figure}

The mesh quality criterion discussed in Field et al. \cite{field2000qualitative} often includes the ratio of the radius of the largest inscribed circle to its diameter, denoted by
\begin{equation}
q=(a+b-c)(a+c-b)(b+c-a)/abc,
\label{q}
\end{equation}
where $a$, $b$, and $c$ are the length of mesh edges. Examples of the $q$ values for various triangle shapes are depicted in Fig. \ref{fig:4.9}. An equilateral triangle has a $q$ value of 1, while a degenerate triangle (with zero area) has a $q$ value of 0. Generally, a triangular mesh group is considered satisfactory if all triangles have $q$ values greater than 0.5 \cite{persson2004simple}.

The superiority of our proposed approach over the classic method is illustrated in Fig. \ref{fig:4.8} (b), where the quality of triangular meshes generated by different methods is compared. The result demonstrates the effectiveness of our approach in improving the quality of input meshes. At the same time, the classic method not only fails to do so but amplifies the proportion of meshes with poor quality.

\begin{figure}[tb]
    \centering
\includegraphics[width=\linewidth]{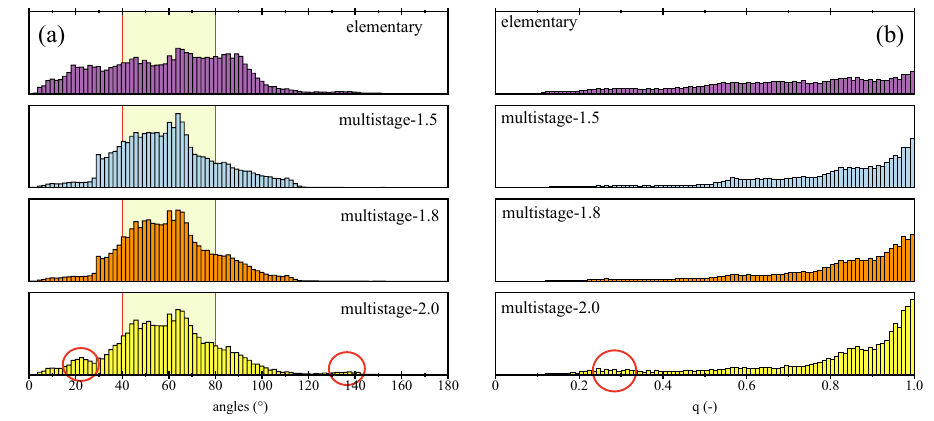}
    \caption{(a) Comparison in angle distributions, (b) Comparison in mesh quality distributions.}
    \label{fig:4.10}
\end{figure}

The angle and quality distributions of mesh groups generated from the elementary novel and multistage subdivisions are presented in Fig. \ref{fig:4.10} (a). The results demonstrate that the multistage subdivisions are significantly better at reducing the proportion of excessively acute or nearly right angles than the elementary novel subdivision. In most cases, it is difficult to observe the differences in the angle distribution among different multistage subdivisions. However, it is worth noting that as the value of $f$ increases, more angles fall within the desired range. In contrast, a notable increase in obtuse and highly acute angles is observed (highlighted by red circles in the figure). The quality distributions are illustrated in Fig. \ref{fig:4.10} (b). Our results demonstrate that the meshes generated through multistage subdivisions exhibit a superior quality distribution compared to those generated through the elementary novel subdivision. However, similarly to the angle distributions, as the fold factor $f$ increases, the proportion of favorably high-quality and some poor-quality meshes (indicated by red circle) increases. The increase of low-quality elements in both mesh qualities and angles is because fewer subdivision stages are applied when the value of $f$ is higher, resulting in fewer geometric optimizations brought by the novel subdivision being implemented.

\begin{figure}[tb]
    \centering
\includegraphics[width=\linewidth]{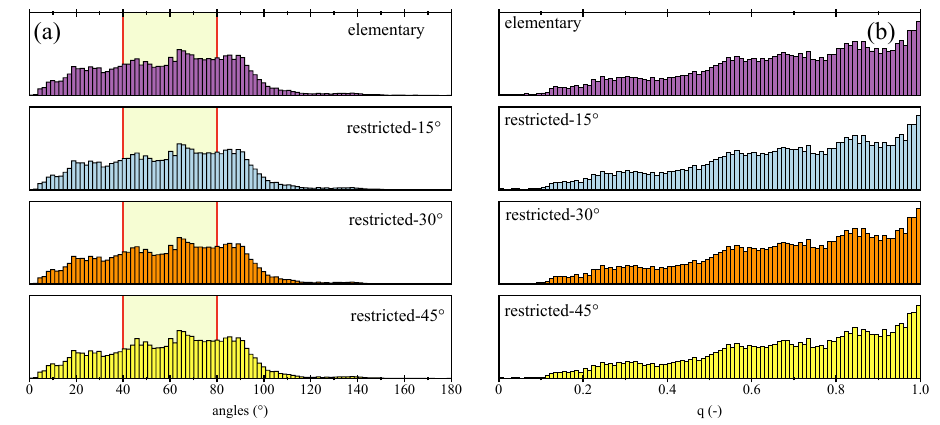}
    \caption{(a) Comparison in angle distributions, (b) Comparison in mesh quality distributions.}
    \label{fig:4.11}
\end{figure}

The angle distributions and quality distributions of mesh groups produced from elementary novel and angle-restricted subdivisions are presented in Fig. \ref{fig:4.11}. In most of the cases, differences in both angle and quality distributions are less apparent. However, the data in Table \ref{tab} shows that the angle-restricted subdivision produces a better angle and quality distribution than the elementary novel subdivision. Moreover, in the scope of this work, the larger the threshold angle $\theta_0$, the more significant the mesh-enhancing capacity becomes.

\section{Conclusion}

This study introduces a novel edge-length-based linear local subdivision strategy for triangular meshes, which can be further enhanced by incorporating two auxiliary techniques.  The investigation results demonstrate several important conclusions.  (1) Our proposed method outperforms the classic method regarding computational performance, element generation, and mesh-enhancing capacity (2) The multistage subdivision demonstrates superior performance in terms of computational time and mesh quality enhancement compared to the elementary novel method. While the fold factor $f$ can provide varying degrees of improvement, a larger value of $f$ generally leads to more favorable computational improvements but may increase the proportion of degenerate meshes. (3) The angle-restricted subdivision has a better computational performance and mesh quality enhancement capability than the elementary method. These differences are generally not noticeable on an exponential axis and monotonically increase as the threshold angle increases within the range of this work. Overall, our proposed strategy, along with its auxiliary schemes, presents a promising approach to improve the computational performance of the subdivision process and any subsequent computing processes while enhancing the quality of generated meshes.

\addtolength{\textheight}{-9cm} 


\bibliographystyle{IEEEtran}

\begin{thebibliography}{10}
\providecommand{\url}[1]{#1}
\csname url@rmstyle\endcsname
\providecommand{\newblock}{\relax}
\providecommand{\bibinfo}[2]{#2}
\providecommand\BIBentrySTDinterwordspacing{\spaceskip=0pt\relax}
\providecommand\BIBentryALTinterwordstretchfactor{4}
\providecommand\BIBentryALTinterwordspacing{\spaceskip=\fontdimen2\font plus
\BIBentryALTinterwordstretchfactor\fontdimen3\font minus
  \fontdimen4\font\relax}
\providecommand\BIBforeignlanguage[2]{{%
\expandafter\ifx\csname l@#1\endcsname\relax
\typeout{** WARNING: IEEEtran.bst: No hyphenation pattern has been}%
\typeout{** loaded for the language `#1'. Using the pattern for}%
\typeout{** the default language instead.}%
\else
\language=\csname l@#1\endcsname
\fi
#2}}

\bibitem{zhang2010spectral}
H.~Zhang, O.~Van~Kaick, and R.~Dyer, ``Spectral mesh processing,'' in
  \emph{Computer graphics forum}, vol.~29, no.~6.\hskip 1em plus 0.5em minus
  0.4em\relax Wiley Online Library, 2010, pp. 1865--1894.

\bibitem{sorkine2006differential}
O.~Sorkine, ``Differential representations for mesh processing,'' in
  \emph{Computer Graphics Forum}, vol.~25, no.~4.\hskip 1em plus 0.5em minus
  0.4em\relax Wiley Online Library, 2006, pp. 789--807.

\bibitem{he2014review}
X.~He, F.~Gu, and A.~Ball, ``A review of numerical analysis of friction stir
  welding,'' \emph{Progress in Materials Science}, vol.~65, pp. 1--66, 2014.

\bibitem{sosnowski2018polyhedral}
M.~Sosnowski, J.~Krzywanski, K.~Grabowska, and R.~Gnatowska, ``Polyhedral
  meshing in numerical analysis of conjugate heat transfer,'' in \emph{EPJ Web
  of Conferences}, vol. 180.\hskip 1em plus 0.5em minus 0.4em\relax EDP
  Sciences, 2018, p. 02096.

\bibitem{petres2007path}
C.~Petres, Y.~Pailhas, P.~Patron, Y.~Petillot, J.~Evans, and D.~Lane, ``Path
  planning for autonomous underwater vehicles,'' \emph{IEEE Transactions on
  Robotics}, vol.~23, no.~2, pp. 331--341, 2007.

\bibitem{bircher2018receding}
A.~Bircher, M.~Kamel, K.~Alexis, H.~Oleynikova, and R.~Siegwart, ``Receding
  horizon path planning for 3d exploration and surface inspection,''
  \emph{Autonomous Robots}, vol.~42, pp. 291--306, 2018.

\bibitem{raj2000modeling}
K.~H. Raj, R.~S. Sharma, S.~Srivastava, and C.~Patvardhan, ``Modeling of
  manufacturing processes with anns for intelligent manufacturing,''
  \emph{International Journal of Machine Tools and Manufacture}, vol.~40,
  no.~6, pp. 851--868, 2000.

\bibitem{zhao2020mixed}
D.~Zhao and W.~Guo, ``Mixed-layer adaptive slicing for robotic additive
  manufacturing (am) based on decomposing and regrouping,'' \emph{Journal of
  Intelligent Manufacturing}, vol.~31, no.~4, pp. 985--1002, 2020.

\bibitem{bolz2002rapid}
J.~Bolz and P.~Schr{\"o}der, ``Rapid evaluation of catmull-clark subdivision
  surfaces,'' in \emph{Proceedings of the seventh international conference on
  3D Web technology}, 2002, pp. 11--17.

\bibitem{guo2019automatic}
J.~Guo, F.~Ding, X.~Jia, and D.-M. Yan, ``Automatic and high-quality surface
  mesh generation for cad models,'' \emph{Computer-Aided Design}, vol. 109, pp.
  49--59, 2019.

\bibitem{alliez2003isotropic}
P.~Alliez, E.~C. De~Verdire, O.~Devillers, and M.~Isenburg, ``Isotropic surface
  remeshing,'' in \emph{2003 Shape Modeling International.}\hskip 1em plus
  0.5em minus 0.4em\relax IEEE, 2003, pp. 49--58.

\bibitem{wu1997advantages}
J.-Y. Wu and R.~Lee, ``The advantages of triangular and tetrahedral edge
  elements for electromagnetic modeling with the finite-element method,''
  \emph{IEEE Transactions on Antennas and Propagation}, vol.~45, no.~9, pp.
  1431--1437, 1997.

\bibitem{lindquist2005comparison}
D.~R. Lindquist and M.~B. Gilest, ``A comparison of numerical schemes on
  triangular and quadrilateral meshes,'' in \emph{11th International Conference
  on Numerical Methods in Fluid Dynamics}.\hskip 1em plus 0.5em minus
  0.4em\relax Springer, 2005, pp. 369--373.

\bibitem{jain2017numerical}
R.~Jain, S.~K. Pal, and S.~B. Singh, ``Numerical modeling methodologies for
  friction stir welding process,'' in \emph{Computational Methods and
  Production Engineering}.\hskip 1em plus 0.5em minus 0.4em\relax Elsevier,
  2017, pp. 125--169.

\bibitem{chen2002automated}
H.~Chen, W.~Sheng, N.~Xi, M.~Song, and Y.~Chen, ``Automated robot trajectory
  planning for spray painting of free-form surfaces in automotive
  manufacturing,'' in \emph{Proceedings 2002 IEEE International Conference on
  Robotics and Automation (Cat. No. 02CH37292)}, vol.~1.\hskip 1em plus 0.5em
  minus 0.4em\relax IEEE, 2002, pp. 450--455.

\bibitem{du2009mesh}
Q.~Du, D.~Wang, and L.~Zhu, ``On mesh geometry and stiffness matrix
  conditioning for general finite element spaces,'' \emph{SIAM journal on
  numerical analysis}, vol.~47, no.~2, pp. 1421--1444, 2009.

\bibitem{beall2003accessing}
M.~W. Beall, J.~Walsh, and M.~S. Shephard, ``Accessing cad geometry for mesh
  generation.'' in \emph{Imr}, 2003, pp. 33--42.

\bibitem{bern1995mesh}
M.~Bern and D.~Eppstein, ``Mesh generation and optimal triangulation,'' in
  \emph{Computing in Euclidean geometry}.\hskip 1em plus 0.5em minus
  0.4em\relax World Scientific, 1995, pp. 47--123.

\bibitem{shimada2011current}
K.~Shimada, ``Current issues and trends in meshing and geometric processing for
  computational engineering analyses,'' 2011.

\bibitem{foucault2008adaptation}
G.~Foucault, J.-C. Cuilli{\`e}re, V.~Fran{\c{c}}ois, J.-C. L{\'e}on, and
  R.~Maranzana, ``Adaptation of cad model topology for finite element
  analysis,'' \emph{Computer-Aided Design}, vol.~40, no.~2, pp. 176--196, 2008.

\bibitem{jamieson1995direct}
R.~Jamieson and H.~Hacker, ``Direct slicing of cad models for rapid
  prototyping,'' \emph{Rapid Prototyping Journal}, vol.~1, no.~2, pp. 4--12,
  1995.

\bibitem{ledalla2018performance}
S.~R.~K. Ledalla, B.~Tirupathi, and V.~Sriram, ``Performance evaluation of
  various stl file mesh refining algorithms applied for fdm-rp process,''
  \emph{Journal of The Institution of Engineers (India): Series C}, vol.~99,
  pp. 339--346, 2018.

\bibitem{king2006cad}
M.~L. King, M.~J. Fisher, and C.~G. Jensen, ``A cad-centric approach to cfd
  analysis with discrete features,'' \emph{Computer-Aided Design and
  Applications}, vol.~3, no. 1-4, pp. 279--288, 2006.

\bibitem{flynn1989cad}
P.~J. Flynn and A.~K. Jain, ``Cad-based computer vision: from cad models to
  relational graphs,'' in \emph{Conference Proceedings., IEEE International
  Conference on Systems, Man and Cybernetics}.\hskip 1em plus 0.5em minus
  0.4em\relax IEEE, 1989, pp. 162--167.

\bibitem{schoberl1997netgen}
J.~Sch{\"o}berl, ``Netgen an advancing front 2d/3d-mesh generator based on
  abstract rules,'' \emph{Computing and visualization in science}, vol.~1,
  no.~1, pp. 41--52, 1997.

\bibitem{geuzaine2009gmsh}
C.~Geuzaine and J.-F. Remacle, ``Gmsh: A 3-d finite element mesh generator with
  built-in pre-and post-processing facilities,'' \emph{International journal
  for numerical methods in engineering}, vol.~79, no.~11, pp. 1309--1331, 2009.

\bibitem{sunil2008automatic}
V.~Sunil and S.~Pande, ``Automatic recognition of features from freeform
  surface cad models,'' \emph{Computer-Aided Design}, vol.~40, no.~4, pp.
  502--517, 2008.

\bibitem{severn2006fast}
A.~Severn and F.~Samavati, ``Fast intersections for subdivision surfaces,'' in
  \emph{International Conference on Computational Science and Its
  Applications}.\hskip 1em plus 0.5em minus 0.4em\relax Springer, 2006, pp.
  91--100.

\bibitem{hoppe1993mesh}
H.~Hoppe, T.~DeRose, T.~Duchamp, J.~McDonald, and W.~Stuetzle, ``Mesh
  optimization,'' in \emph{Proceedings of the 20th annual conference on
  Computer graphics and interactive techniques}, 1993, pp. 19--26.

\bibitem{cignoni1998comparison}
P.~Cignoni, C.~Montani, and R.~Scopigno, ``A comparison of mesh simplification
  algorithms,'' \emph{Computers \& Graphics}, vol.~22, no.~1, pp. 37--54, 1998.

\bibitem{alliez2008recent}
P.~Alliez, G.~Ucelli, C.~Gotsman, and M.~Attene, ``Recent advances in remeshing
  of surfaces,'' \emph{Shape analysis and structuring}, pp. 53--82, 2008.

\bibitem{noii2020adaptive}
N.~Noii, F.~Aldakheel, T.~Wick, and P.~Wriggers, ``An adaptive global--local
  approach for phase-field modeling of anisotropic brittle fracture,''
  \emph{Computer Methods in Applied Mechanics and Engineering}, vol. 361, p.
  112744, 2020.

\bibitem{catmull1998recursively}
E.~Catmull and J.~Clark, ``Recursively generated b-spline surfaces on arbitrary
  topological meshes,'' in \emph{Seminal graphics: pioneering efforts that
  shaped the field}, 1998, pp. 183--188.

\bibitem{rypl2006generation}
D.~Rypl and Z.~Bittnar, ``Generation of computational surface meshes of stl
  models,'' \emph{Journal of Computational and Applied Mathematics}, vol. 192,
  no.~1, pp. 148--151, 2006.

\bibitem{loop1987smooth}
C.~Loop, ``Smooth subdivision surfaces based on triangles,'' 1987.

\bibitem{jacquemin2023smart}
T.~Jacquemin, P.~Suchde, and S.~P. Bordas, ``Smart cloud collocation:
  geometry-aware adaptivity directly from cad,'' \emph{Computer-Aided Design},
  vol. 154, p. 103409, 2023.

\bibitem{surazhsky2003isotropic}
V.~Surazhsky, P.~Alliez, and C.~Gotsman, ``Isotropic remeshing of surfaces: a
  local parameterization approach,'' Ph.D. dissertation, INRIA, 2003.

\bibitem{sullivan2019pyvista}
C.~Sullivan and A.~Kaszynski, ``Pyvista: 3d plotting and mesh analysis through
  a streamlined interface for the visualization toolkit (vtk),'' \emph{Journal
  of Open Source Software}, vol.~4, no.~37, p. 1450, 2019.

\bibitem{bank1983some}
R.~E. Bank, A.~H. Sherman, and A.~Weiser, ``Some refinement algorithms and data
  structures for regular local mesh refinement,'' \emph{Scientific Computing,
  Applications of Mathematics and Computing to the Physical Sciences}, vol.~1,
  pp. 3--17, 1983.

\bibitem{bern2000mesh}
M.~W. Bern and P.~E. Plassmann, ``Mesh generation.'' \emph{Handbook of
  computational geometry}, vol.~38, 2000.

\bibitem{muller2000subdivision}
K.~M{\"u}ller and S.~Havemann, ``Subdivision surface tesselation on the fly
  using a versatile mesh data structure,'' in \emph{Computer Graphics Forum},
  vol.~19, no.~3.\hskip 1em plus 0.5em minus 0.4em\relax Wiley Online Library,
  2000, pp. 151--159.

\bibitem{pulli1996fast}
K.~Pulli and M.~Segal, ``Fast rendering of subdivision surfaces,'' in
  \emph{Rendering Techniques’ 96: Proceedings of the Eurographics Workshop in
  Porto, Portugal, June 17--19, 1996 7}.\hskip 1em plus 0.5em minus 0.4em\relax
  Springer, 1996, pp. 61--70.

\bibitem{field2000qualitative}
D.~A. Field, ``Qualitative measures for initial meshes,'' \emph{International
  Journal for Numerical Methods in Engineering}, vol.~47, no.~4, pp. 887--906,
  2000.

\bibitem{persson2004simple}
P.-O. Persson and G.~Strang, ``A simple mesh generator in matlab,'' \emph{SIAM
  review}, vol.~46, no.~2, pp. 329--345, 2004.

\end{thebibliography}

\end{document}